\documentclass[%
reprint,
 amsmath,amssymb,
 aps,
]{revtex4-1}

\usepackage{graphicx}
\graphicspath{
{./}
{/home/jripley/SphericalCollapse/EdGB_Exp/plt_files/}
{/home/jripley/SphericalCollapse/EdGB_Exp/plt_files_for_paper/}
{/home/jripley/SphericalCollapse/EdGB_Exp/plt_files_for_paper/4097_8193_4_2.5e-01_1.0e-12_100.0_NewtonBanded_relaxationFullMatrix_2by2_0_100.0000_r2Gaussian_2.0000000000e-02_10.00_30.00_ingoing/}
}
\usepackage{bm}
\usepackage{color}

\begin{document}


\title{Hyperbolicity in Spherical Gravitational Collapse in a Horndeski Theory}

\author{Justin L. Ripley}
\email{jripley@princeton.edu}

\author{Frans Pretorius}%
\email{fpretori@princeton.edu}

\affiliation{%
 Department of Physics, Princeton University, Princeton, New Jersey 08544, USA.
}%


\date{\today}

\begin{abstract}
	We numerically study spherical gravitational collapse in
shift symmetric Einstein dilaton Gauss Bonnet (EdGB)
gravity. We find evidence that there are open sets of initial data for
which the character of the system of equations changes from hyperbolic
to elliptic type in a compact region of the spacetime. In these cases
evolution of the system, treated as a hyperbolic initial boundary value
problem, leads to the equations of motion becoming ill-posed when
the elliptic region forms.
No singularities or discontinuities are encountered on the corresponding
effective ``Cauchy horizon". Therefore it is conceivable that a
well-posed formulation of EdGB gravity (at least within spherical symmetry)
may be possible if the equations are appropriately treated as mixed type. 
 
\end{abstract}

\maketitle


\section{\label{Sec:Introduction}Introduction}
	Gravitational wave observation of the binary inspiral
of compact objects provides an
opportunity to study and test general relativity (GR) in the strong field,
dynamical regime. Ideal for such tests are binary black hole (BH) mergers,
due to the uniqueness properties of BHs in GR, and that ostensibly
their circumbinary environments are sufficiently free of matter to not
affect the waveform at a (presently) measurable level. One problem
achieving the best possible constraints from
current data~\cite{LIGOScientific:2018mvr}
is the dearth of interesting, viable alternatives to GR that can 
make concrete predictions in the late inspiral and merger regime
(see e.g.~\cite{Yunes:2016jcc}). Specific to this context, 
by {\em interesting}, we mean theories that are consistent with GR
in the well-tested weak field regime, yet still predict significant
differences for the mergers of BHs; by {\em viable}, we mean theories
that offer a classically well-posed initial value problem, requisite
for computing waveforms to confront with data.

A common scheme to designing modified gravity theories is to add
curvature scalars beyond the Ricci scalar $R$ to the Einstein-Hilbert action. 
The problem with this is typically the resultant classical equations
of motion are partial differential equations (PDEs) with higher
than second-order 
derivatives, generically making them Ostrogradsky unstable
\cite{Woodard:2006nt,Woodard:2015zca}. Two approaches are being pursued
to cure this: (1) treating GR as an effective field theory
(EFT) and calculating perturbative
corrections from the higher order terms
\cite{Okounkova:2017yby,Okounkova:2018pql,
Benkel:2016kcq,Witek:2018dmd,Allwright:2018rut},
and (2) modifying the problematic terms in a somewhat
ad-hoc fashion inspired by the Israel-Stewart prescription for making
relativistic, viscous hydrodynamics well-posed~\cite{Cayuso:2017iqc}.
However, there are sub-classes of higher curvature modified gravity theories
that still only have second order PDEs, offering the hope that
their full, classical equations of motion are well-posed without
the need to resort to the above approximation schemes.
One such theory is 
Einstein dilaton Gauss-Bonnet (EdGB) gravity
(see e.g. \cite{Kanti:1995vq,Maeda:2009uy,Yagi:2015oca}
and the references therein), which is the focus
of this paper. In particular, we restrict to a version with
linear coupling between the scalar field $\phi$ and curvature,
which is the simplest member of the shift symmetric class of EdGB theories.

What is interesting about this EdGB gravity theory
in the sense of the word discussed above,
is that it does not admit the Schwarzschild or Kerr BH solutions
of GR. Instead, the analogue BH solutions only exist above
a minimum length scale related to the coupling constant $\lambda$
in the theory,
and feature scalar
``hair''~\cite{Kanti:1995vq,Sotiriou:2013qea,Sotiriou:2014pfa}.
Moreover, for values of $\lambda$ that would produce significant changes
in stellar mass BHs, the corresponding effect on material compact objects
such as neutron stars is insignificant~\cite{Yagi:2015oca}, implying
this theory
could be consistent with current GR tests, yet give different results for
stellar mass BH mergers.
However, it is still unknown under what conditions, if any, EdGB
gravity is viable.
EdGB gravity can be considered a member of the Horndeski class
of scalar-tensor 
theories~\cite{Horndeski1974}, though the mapping between the latter form
and the one we use here where the Gauss-Bonnet scalar $\mathcal{G}$
is explicit in 
the action is highly non-trivial~\cite{Kobayashi:2011nu}.
The only systematic study of the well-posedness of EdGB gravity we are aware 
of~\cite{Papallo:2017qvl,Papallo:2017ddx}
considered the linearized equations in Horndeski form in the small
coupling parameter limit,
and found that generically the PDEs are only weakly hyperbolic
within a certain class of ``generalized harmonic'' gauges. 

In this work we initiate a study of the hyperbolic properties of
4-dimensional EdGB
gravity using numerical solutions of the full equations in the strong field, 
dynamical regime. As mentioned above, one of our main goals is to
ascertain whether EdGB gravity can be relevant for (or constrained with)
observable BH mergers. Therefore as a first step toward this goal
we restrict to spherically symmetric, asymptotically flat spacetimes,
as this is the simplest symmetry reduced setting allowing BH solutions.
As in GR, there are no gravitational
waves then, and all the dynamics are driven by the scalar field.   
We thus cannot yet begin to fully address whether BH merger spacetimes
are sufficiently ``non-generic'' to avoid the conclusions reached
in the above mentioned studies~\cite{Papallo:2017qvl,Papallo:2017ddx} (assuming 
weak hyperbolicity cannot be lifted to strong hyperbolicity 
by some novel gauge condition). What we can learn from this study
are which scenarios are free of pathologies in spherical symmetry,
and use this to guide future studies were we will relax this symmetry restriction.

Our initial survey focused on several regimes, including the weak
field/weak coupling limit
where an initial concentration of scalar field energy eventually disperses
beyond the integration domain, and the 
strong field/weak coupling regime where the scalar field begins to
collapse to a BH 
(though since our present code uses Schwarzschild-like coordinates
we cannot evolve 
beyond BH formation). In all these cases,
which we will discuss in detail in an upcoming
paper~\cite{RipleyPretorius:2019},
we see no break down in hyperbolicity over the integration domain. 
Here we present novel results on evolution in the strong field/strong coupling
regime, though well below the threshold of black hole formation. In contrast to
the weak coupling cases, we find a swath of parameter space where evolution
leads to a region of spacetime where the PDEs switch character from
hyperbolic to elliptic, implying here the PDEs are actually what is referred to as {\em mixed type}.
Treated as an initial value problem (IVP), the boundary of the elliptic region
is thus effectively a ``Cauchy horizon'', beyond which the equations
become ill-posed
as an IVP (though this is not a Cauchy horizon in the usual sense
it is employed
in GR, hence the quotes, as the boundary is not null, and from the
perspective of the metric 
the interior and exterior regions are still in causal contact with
one another).

	Regarding previous work on the
dynamical loss of hyperbolicity in solutions of Horndeski theories,
this behavior has also been identified 
at the perturbative level in many Horndeski theories
applied to cosmological scenarios
(see e.g.\cite{Quiros:2019ktw,Kobayashi:2019hrl}
and the references therein),
though there what we call the elliptic region is usually
described as a place where the theory begins to suffer from a 
``Laplacian'' or ``gradient'' instability. This is a bit of a misnomer
as these labels imply some form of physical instability.
For with these ``instabilities''
there either is no problem if an appropriate, physically
well-motivated mixed type formulation
can be devised, or, as pointed out
by \cite{Reall:2014pwa,Papallo:2015rna}
(see also, e.g. \cite{Papallo:2017qvl,Papallo:2017ddx,Ijjas:2018cdm})
the problem is much worse than a physical instability if we demand that the
only physically sensible classical theories are those where predictions must
be made in the sense of a well-posed IVP. 
Under the latter condition, initial data within an IVP formulation
that leads to an elliptic region would either need to be
excluded, or the formation of the elliptic region would need to be
regarded as signalling the break down of the theory
(as for example singularity formation is in GR).

	Researchers investigating the
hyperbolicity of higher dimensional Lovelock theories 
have noted the existence of elliptic regions in black
hole and cosmological solutions to those theories.
Dimensionally reduced to 4 spacetime dimensions these
Lovelock theories can resemble EdGB gravity (see e.g.\cite{Charmousis:2014mia}).
The authors in \cite{Reall:2014pwa} 
find that the equations of motion for linear gravitational perturbations
of sufficiently small static, spherically symmetric
Lovelock black holes are no longer hyperbolic in a
region outside the black hole horizon. This may be related
to the fact discussed above that EdGB gravity does not have
static BH solutions below a certain scale.
The authors in
\cite{Papallo:2017qvl} show that the dynamical violation of
hyperbolicity in Lovelock theories applied to cosmological evolution could
in principle occur, although
they do not construct explicit solutions where initially hyperbolic
equations evolve to form an elliptic region. 

An outline of the rest of this paper is as follows.
In Sec.~\ref{Sec:EquationsOfMotion} we present the equations of motion of
EdGB gravity, along with our choice of coordinates and variables, and the form
of the equations we ultimately solve in the code.
In Sec.~\ref{sec:characteristics} we outline the calculation
of the characteristics of the theory in spherical symmetry.
Sec.~\ref{sec:simulations} contains results on the class of initial
data where evolution can lead to break down of hyperbolicity.
We conclude in Sec.~\ref{sec:Conclusion} with a discussion of
the implications of this for EdGB gravity.
We briefly describe our numerical methods and a convergence
study in the Appendix, leaving more details and
results from different parts of parameter space to~\cite{RipleyPretorius:2019}.
We work in geometric units and use MTW \cite{misner1973gravitation} 
sign conventions for the metric tensor, etc.

\section{\label{Sec:EquationsOfMotion}Equations of motion}
We consider the following EdGB action
\begin{align}
\label{eq:sGB_action}
	S = \frac{1}{2}
	\int d^4x\sqrt{-g} 
	\left(R - (\nabla\phi)^2 
	 + 2\lambda\phi \mathcal{G} 
	\right).	
\end{align}
	The Gauss-Bonnet scalar can be written in terms of the Riemann tensor as  
$	\mathcal{G}	
	\equiv
	\frac{1}{4}
	\delta^{\mu\nu\alpha\beta}_{\lambda\sigma\gamma\delta}
	R^{\lambda\sigma}{}_{\mu\nu}R^{\gamma\delta}{}_{\alpha\beta}
$, where $\delta^{\mu\nu\alpha\beta}_{\lambda\sigma\gamma\delta}$ is the
generalized Kronecker delta.
Varying \eqref{eq:sGB_action} in turn with respect to the metric
$g^{\mu\nu}$ and 
scalar $\phi$ gives
\begin{subequations}\label{eq_EinsteinEquations:main}
\begin{align}
\label{eq_EinsteinEquations:metriceom}
	R_{\mu\nu} - \frac{1}{2}g_{\mu\nu}R
	+2\lambda \delta^{\gamma\delta\kappa\lambda}_{\alpha\beta\rho\sigma}
	R^{\rho\sigma}{}_{\kappa\lambda}
	\left(\nabla^{\alpha}\nabla_{\gamma}\phi\right)
	\delta^\beta{}_{(\mu} g_{\nu)\delta}
	\nonumber \\
	- \nabla_{\mu}\phi\nabla_{\nu}\phi + \frac{1}{2}g_{\mu\nu}(\nabla\phi)^2
	= 0
	, 
\end{align}
\begin{align}
\label{eq_EinsteinEquations:phieom}
	\Box\phi + \lambda \mathcal{G} 
	= 0
	. 
\end{align}
\end{subequations}
	We choose to write the line element in the form 
\begin{align}
\label{eq:polar_coordinates}
	ds^2=-e^{2A(t,r)}dt^2 + e^{2B(t,r)}dr^2 
	+ r^2\left(d\vartheta^2+\mathrm{sin}^2\vartheta d\varphi^2\right)
	.
\end{align}
Defining the variables 
$	Q(t,r)
	\equiv 
	\partial_r\phi
$ and
$	P(t,r)
	\equiv 
	e^{-A+B}\partial_t\phi
$ ,
and taking appropriate algebraic combinations of
\eqref{eq_EinsteinEquations:phieom} and the non-trivial
components of ~\eqref{eq_EinsteinEquations:metriceom} results in 
the following system of PDEs: 
\begin{widetext}
\begin{subequations}\label{eq_sGB_solved:main}
\begin{align}
\label{eq_sGB_solved:rDer_A}
	& 
	\Bigg(
		\mathcal{I}^2
	- 	32\lambda^2\mathcal{B}^2
	+ 	128 \lambda^2e^{-2B}\mathcal{B}
		\left(
		1-2\lambda\left(3e^{-2B}+1\right)\frac{Q}{r}
		\right)\frac{\partial_rB}{r}
	+ 256\lambda^3\mathcal{B}^2\left(
		e^{-2B}\partial_rQ 
		- e^{-B} r P\mathcal{K}
		\right)
	\Bigg)\partial_rA 
	\nonumber \\
	+ & 4 \lambda e^{-3B}\mathcal{B}
	\left(
		128\lambda^2e^{2B}r\mathcal{B}P \mathcal{K}
	- 	4 \lambda e^B P^2
	+	e^B\left(r e^{2B}-12\lambda Q\right)Q
	\right)\partial_rB
	- 512\lambda^3r e^{-B}\mathcal{B}^2
		\mathcal{K}\partial_rP
	\nonumber \\
	- &	4\lambda r\mathcal{B}\mathcal{I}
		\partial_rQ
	- \frac{r\mathcal{B}}{2}
		\left(e^{2B}+128\lambda^2\mathcal{K}^2\right)
	+ 4 \lambda\mathcal{B}\left(-1+128\lambda^2\mathcal{K}^2\right)Q
	+ 2\lambda e^{-2B}Q^3 
	\nonumber \\
	+ & \left(
		64\lambda^2e^{-2B}r\mathcal{B}
	-	16r^3\lambda^2\mathcal{B}^2-\frac{r^3}{4}
	\right)\left(\frac{Q}{r}\right)^2
	+ 4 \lambda r^2e^BP\mathcal{I}\mathcal{B}
		\mathcal{K}
	+ \left(
		16\lambda^2r \mathcal{B}^2
	-	\frac{r}{4}\mathcal{I}
	\right) P^2
	= 0 
	, 
\end{align}
\begin{align}
\label{eq_sGB_solved:rDer_B}
	&
	  \left(1 + 4\lambda\left(1-3e^{-2B}\right)\frac{Q}{r}\right)\partial_rB
	- \frac{r}{4}\left(Q^2+P^2\right)-\frac{1-e^{2B}}{2r}
	+ 4\lambda r\mathcal{B}\left(
	- \partial_rQ
	+ r e^BP \mathcal{K}
	\right)
	= 0
	, 
\end{align}
\begin{align}
\label{eq_sGB_solved:tDer_Q}
	&
	\partial_tQ - \partial_r\left(e^{A-B}P\right)
	= 0 
	, 
\end{align}
\begin{align}
\label{eq_sGB_solved:tDer_P}
	\left(
	\mathcal{I}
	+ 64\lambda^2e^{-2B}\mathcal{B}\frac{\partial_rB}{r} 
	\right)
	\partial_tP 
	- \left(\mathcal{I} 
	-64\lambda^2e^{-2B}\mathcal{B}\frac{\partial_rA}{r}
	\right)
	\frac{1}{r^2}\partial_r\left(r^2e^{A-B}Q\right)
	+ 
		16\lambda e^{A-B}\mathcal{I}
		\left(
		\frac{\partial_rA}{r}\frac{\partial_rB}{r} - \mathcal{K}^2
		\right)
	\nonumber \\
	+ 4\lambda e^{A-B}\mathcal{B}\Bigg(
	  \left(P^2-Q^2\right)
	+ 32\lambda r Q \mathcal{K}^2
	- 16 \lambda e^{-2B}\frac{Q}{r} (\partial_rA)^2
	+ 16 \lambda e^{-B} \left(
			\left(\partial_rB-\partial_rA\right)P - 2\partial_rP
		\right) \mathcal{K}
	\nonumber \\
	+ 2\frac{\partial_rB}{r}
	+ 2 \left(
			-1 - 16\lambda e^{-2B}\frac{Q}{r}
			-2r\left(1-4\lambda e^{-2B}\frac{Q}{r}\right)\partial_rB
		\right)\frac{\partial_rA}{r}
	\Bigg)	
	= 0 
	, 
\end{align}
\end{subequations}
\end{widetext}
	where 
$\mathcal{B}\equiv (1-e^{-2B})/r^2$, 
$\mathcal{I}\equiv 1 - 8\lambda e^{-2B}Q/r$, 
and 
\begin{align}
	\mathcal{K} & \equiv
	e^B\frac{
		\frac{PQ}{2} 
		+ 4 \lambda \mathcal{B}\left(-P\partial_rB+\partial_rP\right)
	}{
	e^{2B} + 4\lambda\left(-3 + e^{2B}\right)\frac{Q}{r}
	}
	.
\end{align}	
Eqs.~\eqref{eq_sGB_solved:main}, though more involved, retain the
basic structure of the spherically symmetric Einstein massless-scalar
system ($\lambda\to0$). Namely, \eqref{eq_sGB_solved:rDer_A} and 
\eqref{eq_sGB_solved:rDer_B} can be considered constraint equations
for the metric variables $A$ and $B$ given data for $P$
and $Q$ on any $t={\rm const.}$
time slice; then 
\eqref{eq_sGB_solved:tDer_Q} and \eqref{eq_sGB_solved:tDer_P} can be considered
evolution equations (where hyperbolic) for $P$ and $Q$. Moreover,
as in GR, the 
system of PDEs (\ref{eq_EinsteinEquations:metriceom}) is
over-determined, and can 
provide an independent evolution equation for one of the metric functions;
we do not solve this equation, rather we monitor
its convergence to zero (or more specifically
its proxy in the $\vartheta\vartheta$
component of (\ref{eq_EinsteinEquations:metriceom}))
as a check for the correctness of our solution. Further details on our setup,
along with convergence results may be found in the Appendix.

\section{\label{sec:characteristics}Characteristics}
We follow standard techniques to compute the characteristic structure
of our system of PDEs
(e.g.\cite{kreiss1989initial,gustafsson1995time,courant1962methods}).
For constructing the principal symbol it suffices to
only consider the $P,Q$ subsystem,
as $A$ and $B$ are fully constrained. We therefore begin
by algebraically
solving for $\partial_rA$, $\partial_rB$
in (\ref{eq_sGB_solved:rDer_A},\ref{eq_sGB_solved:rDer_B})
to write (\ref{eq_sGB_solved:tDer_Q},\ref{eq_sGB_solved:tDer_P})
as a system of equations of the form
\begin{align}
\label{eq_sGB:just_QP}
	E^{(I)}\left[
		A,B,v^{(J)},\partial_rv^{(J)},\partial_tv^{(J)}
	\right]
	& = 0 
	,
\end{align}
	where $I,J$ run over the labels $(1,2)$, and
$v^{(1)}\equiv Q$ and $v^{(2)}\equiv P$.
Introducing the characteristic covector $\xi_a$,
where $a$ runs over the spacetime indices $(t,r)$,
the principal symbol is 
\begin{align}
	\mathfrak{p}_J^I(\xi_a)
	\equiv
	\frac{\delta E^{(I)}}{\delta\left(\partial_av^{(J)}\right)}\xi_a
	.
\end{align}
Solving the characteristic equation 
$
	\mathrm{det}\left[\mathfrak{p}^I_J(\xi_a)\right]=0
$ for the characteristic
covector, we obtain the following
equation for the characteristic speeds $c\equiv -\xi_t/\xi_r$.
\begin{align}
\label{eq:characteristic_speeds}
	c_{\pm} = \frac{1}{2}
	\left(
		 \mathrm{tr}\left[\mathfrak{c}^I_J\right]
	\pm \sqrt{
			\mathrm{tr}\left[\mathfrak{c}^I_J\right]^2 
		-	4\mathrm{det}\left[\mathfrak{c}^I_J\right] 
		}
	\right)
	.
\end{align} 
	where
\begin{align}
	\mathfrak{c}^I_J 
	\equiv
	\left(
		\frac{\delta E^{(I)}}{\delta\left(\partial_tv^{(K)}\right)}
	\right)^{-1}
	\left(
		\frac{\delta E^{(K)}}{\delta\left(\partial_rv^{(J)}\right)}
	\right)
	.
\end{align}
The sign of the discriminant $\mathcal{D}$
of Eq.~\eqref{eq:characteristic_speeds}
determines the character of the dynamical degree of freedom:
when $\mathcal{D}>0$ it is hyperbolic, when $\mathcal{D}=0$ it is
parabolic, and when $\mathcal{D}<0$ it is elliptic
(when hyperbolic, this $2\times2$ principal symbol is
always strongly hyperbolic in the sense of possessing a complete
set of Eigenvectors).
In the limit $\lambda=0$,
the characteristic speeds reduce to those of GR:
$c_{\pm} = \pm e^{A-B}$ and the dynamical degree of freedom is
always hyperbolic (for $A$ and $B$ finite and real). 
\section{\label{sec:simulations}Results}
	We present results from numerical solution
of (\ref{eq_sGB_solved:main}) for the following family of initial data
\begin{align}
\label{eq:initial_data}
	\phi(r)|_{t=0} 
	= 
	a_0 \left(\frac{r}{w_0}\right)^2 
	\mathrm{exp}\left(-\left(\frac{r-r_0}{w_0}\right)^2\right)
	,
\end{align}
giving $Q|_{t=0}=\partial_r\phi|_{t=0}$, and $P$ is chosen
to make the initial pulse be approximately
ingoing: $P|_{t=0}=-(Q+\phi/r|)_{t=0}$;
$a_0,r_0$ and $w_0$ are constants.

To characterize the strength of the EdGB modification, we perform the following
dimensional analysis. For a compact source of scalar field energy
with characteristic length scale $L$, $|\nabla \phi|\sim|\phi_0|/L$, where
$|\phi_0|$ is the maximum difference between $\phi$ and $\phi(t,r=\infty)$
(as we consider a shift symmetric theory).
In GR,
$|R_{\mu\alpha\nu\beta}|\sim m/L^3$, where $m$ is the Arnowitt-Deser-Misner
(ADM) mass. 
Using these expressions to characterize
the magnitudes of the various terms in
(\ref{eq_EinsteinEquations:metriceom}), and
noting that $\lambda$ has dimension $length^2$, we define
a dimensionless parameter
\begin{align}
\label{eq:dimensionlessRatio_lambdaLphi}
	\eta 
	\equiv &
	\frac{\lambda}{L^2}|\phi_0|
	, 
\end{align}
so that for $\eta>1$ we expect strong modifications from GR solutions.
For the class of initial data above (to within factors of a few)
$L\sim w_0$ and 
$|\phi_0|\sim a_0 (r_0/w_0)^2$. 

The gravitational strength of the initial
data can be characterized by the compaction $C\equiv m/L$. 
Here we present results on cases with large
GR-modifications ($\eta\gtrsim 1$), and low ($C\ll 1$) to moderately
strong field,
but not black hole forming ($C\lesssim1$). We first show results
from one typical case, then a survey confirming the scaling 
relation (\ref{eq:dimensionlessRatio_lambdaLphi}) above.

\subsection{An example of loss of hyperbolicity when $\eta\gtrsim 1$}
\begin{figure} 
	\centering 
	\includegraphics[width=1.0\columnwidth]{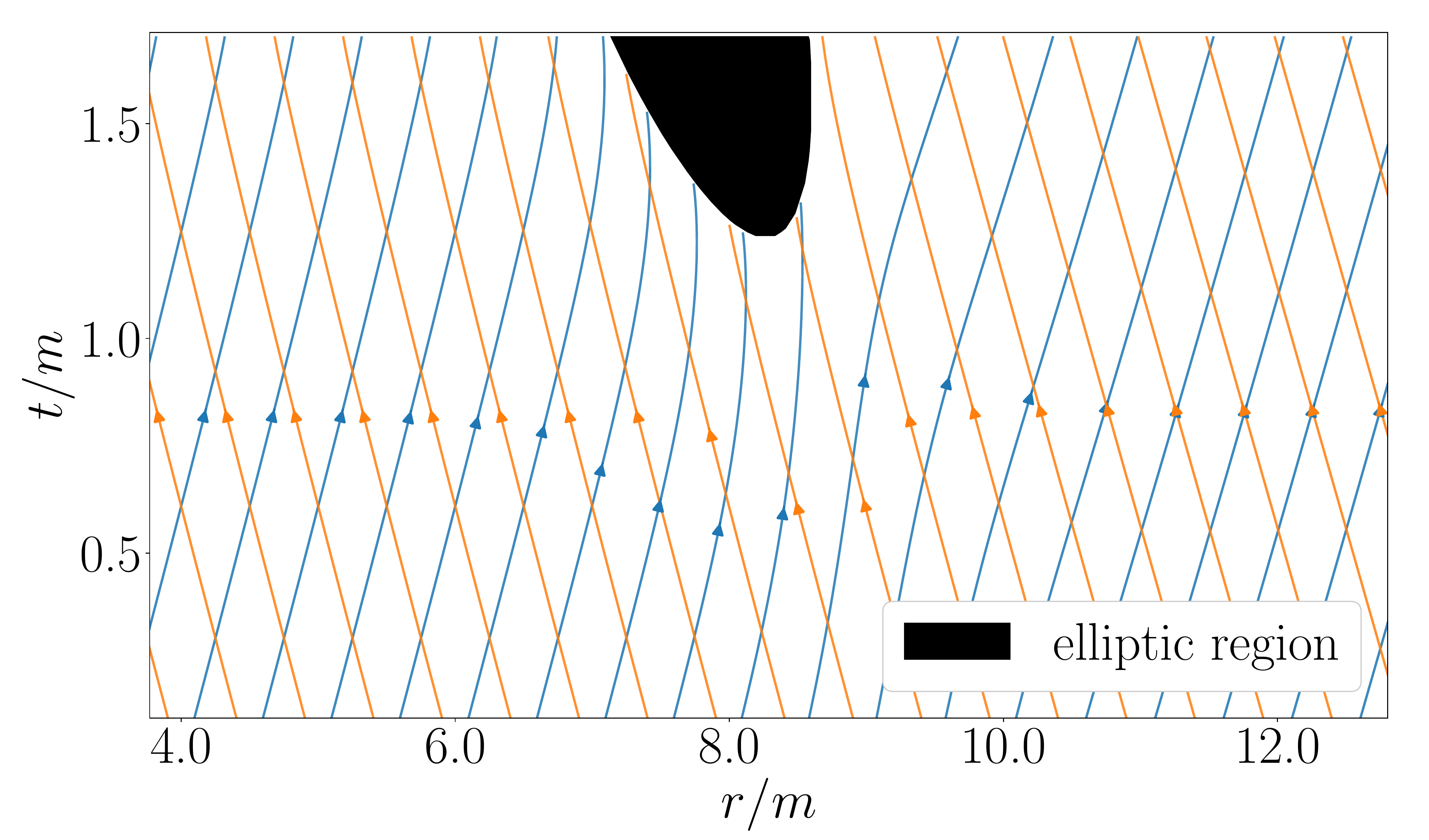}
\caption{Integral curves of the ingoing (red) and
outgoing (blue) characteristic vectors
$(1,c_-)$ and $(1,c_+)$, respectively.}
\label{fig:characteristics}
\end{figure}

\begin{figure} 
\centering
	\includegraphics[width=0.85\columnwidth]{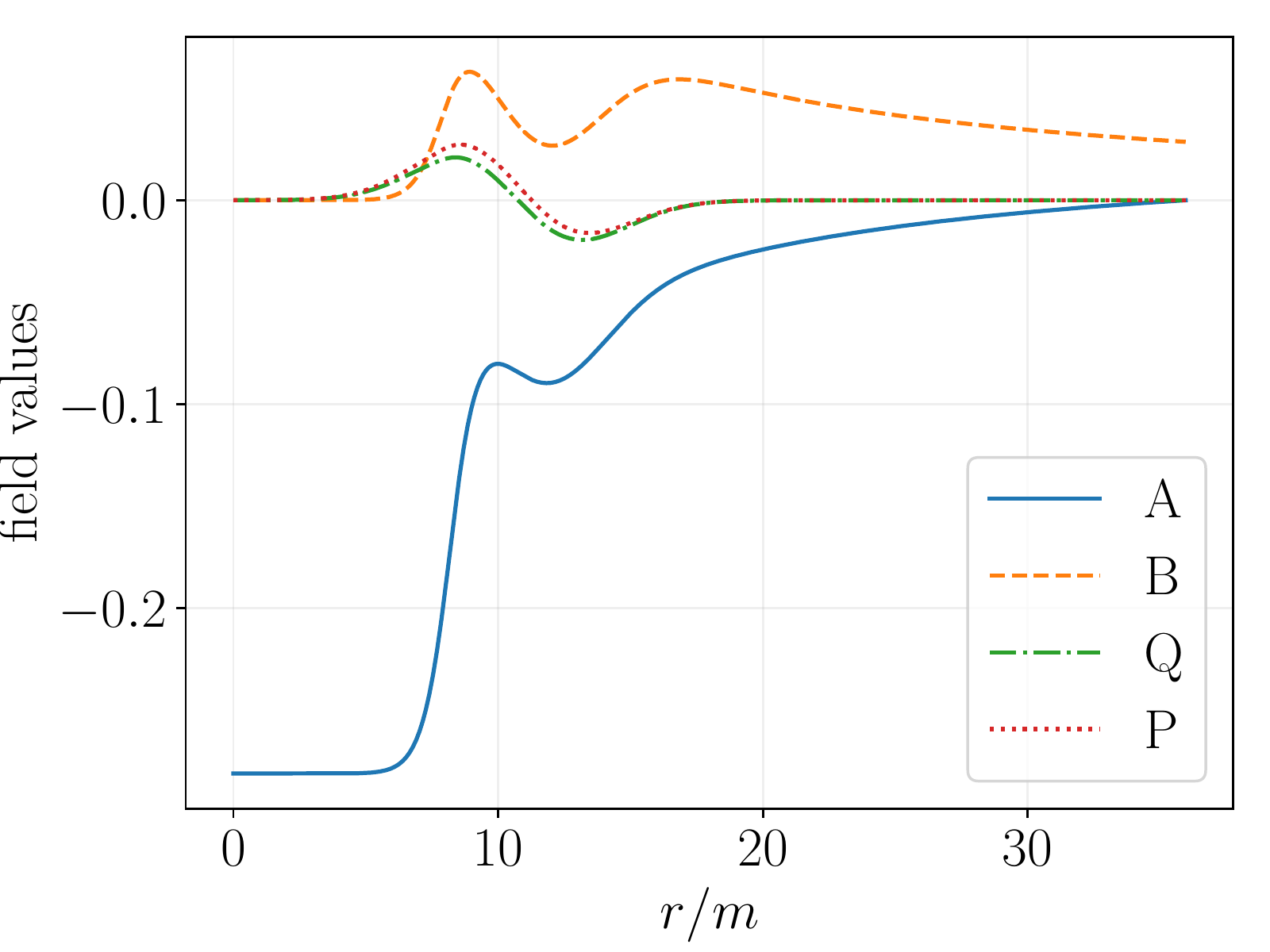}
\caption{
Field variables at the time ($t/m\sim 1.24$)
just before formation of the elliptic region and
subsequent loss of convergence.
}
\label{Fig:field_vals}
\end{figure}

For the specific example, we choose $a_0=0.02$, $r_0=30$, $w_0=10$,
and $\lambda=100$.  The ADM mass is $m\sim2.8$. For this initial data,
the characteristic speeds are initially real, indicating the scalar
field subsystem is hyperbolic. However when $t/m$ reaches $\sim 1.2$, 
a region forms where the characteristic speeds become imaginary, indicating
the character of the equations have become elliptic there---see 
Fig.~\ref{fig:characteristics}. 
Throughout the simulation evolution the \emph{spacetime} outgoing null
characteristic speeds $e^{A-B}$ remain positive
and well away from zero, hence the elliptic region is
not ``censored'' by spacetime
causal structure.
Moreover, at the time the elliptic region first forms, 
all field variables are smooth and finite---see Fig.~\ref{Fig:field_vals}.

\begin{figure}
	\centering
	\includegraphics[width=0.9\columnwidth]{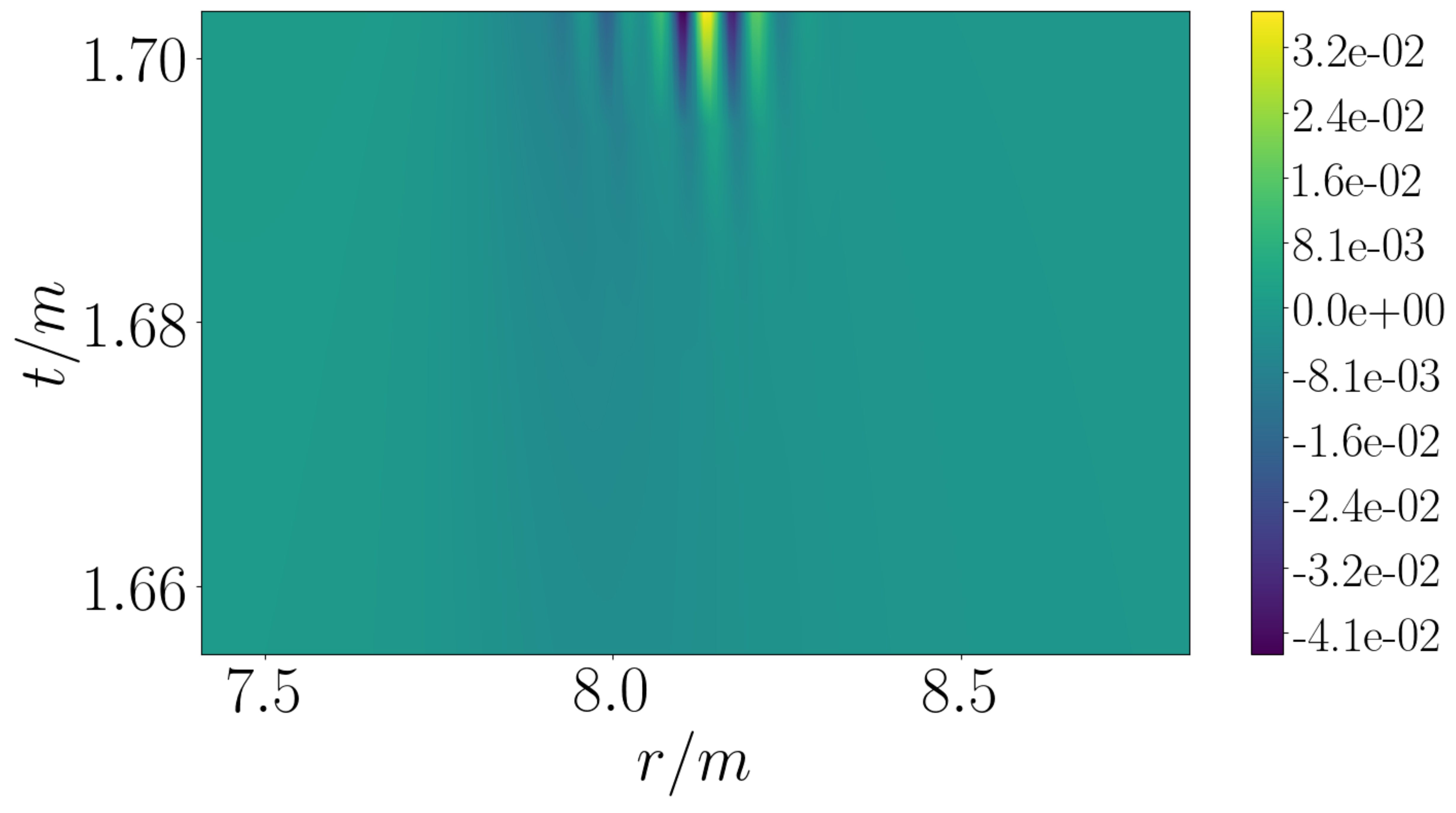}
\caption{Plot of $\partial_r^2P$ inside the elliptic
region just before the code crashes. The oscillations occur within the
elliptic region---compare the axis scales with that of
Fig.~\ref{fig:characteristics}.
}
\label{fig:oscillationsInEllipticRegion}
\end{figure}

	When an elliptic region forms, we expect the PDEs, solved as an IVP,
to become ill-posed. The way this is expected to manifest in a numerical
hyperbolic solution scheme, as we use, is that short wavelength solution
components will begin to grow exponentially, at a rate inversely proportional
to their wavelength.
Since at the analytic level our initial data is
smooth and does not have features smaller than the initial width of
the elliptic
region, the seeds of the growing modes will come from numerical truncation
error. The fastest growing modes are expected to roughly be on
order of the mesh spacing of our finite differencing method.
Fig.~\ref{fig:oscillationsInEllipticRegion}
shows such a growing mode 
just before the code crash at $t/m \sim 1.7$ in this example.

These results imply that the system of PDEs~\eqref{eq_sGB_solved:main}
are of {\em mixed type}, e.g.\cite{otway2015elliptic,rassias1990lecture}.
Such equations can have regions where the PDEs are hyperbolic,
others where they are elliptic, and are parabolic on the co-dimension
one surfaces separating these regions, called {\em sonic lines}.
Given that in Fig.\ref{fig:characteristics} it is evident
that the characteristics do not intersect the sonic line tangentially,
and that the characteristic speeds go to zero on this sonic line,
we surmise that at least qualitatively the EdGB equations in this
scenario are of {\em Tricomi type} (in contrast to {\em Keldysh type}
where the characteristics meet the sonic line tangentially,
and the corresponding speeds diverge there).

Uniqueness results have been obtained for the
Tricomi
equation~\cite{morawetz_dirichlet_problem_tricomi,otway2015elliptic},
and so one could speculate similar results might hold here.
However, given the rather bizarre nature of the boundary conditions needed
for the proof (specification of boundary data on a ``future'' segment
within the elliptic region, and free data along only {\em one}
of the two characteristic surfaces defining the boundary
of the hyperbolic region),
it is unclear how this could be implemented in a manner that is physically
sensible. Moreover, we cannot confidently
claim that the picture we have presented here of the sonic line is
robust, as strictly speaking in the continuum limit we will
only achieve convergence up to the first instant the sonic line is 
encountered. Even restricting to the purely hyperbolic region uncovered
at some fixed resolution, the problem, as mentioned above, is
that the elliptic region is not censored.
Adding any other matter fields to the problem,
or metric perturbations away from spherical symmetry to allow 
for gravitational waves, will causally connect the elliptic
region to the rest of the
domain. Therefore, the effective Cauchy horizon here would not be the
sonic line, but the boundary of the volume of spacetime in causal contact
with the union of elliptic regions.

Another question that arises is to what extent the formation of the elliptic
region is gauge invariant. While characteristics are invariant
under point transformations
(e.g. \cite{courant1962methods}), it is less clear that this must be so
under full spacetime coordinate transformations, even 
restricting to manifestly spherical coordinates where $r$ remains an areal
radius. The problem then is if we consider a new time
coordinate $\bar{t}=\bar{t}(r,t)$,
and wish to pose a Cauchy IVP with respect to $\bar{t}$ (and not simply
transform the solution from $(t,r)$ to $(\bar{t},r)$), we generically
introduce a new gauge degree of freedom (the shift vector $\beta$), and some
new PDE must be specified to solve for it. Generically, such
a prescription will 
mix the metric variables with the scalar degree of freedom in a manner where
we cannot cleanly separate the gravitational gauge degree of freedom
from that of the scalar (or said another way, this will change the
rank and structure
of the principal symbol in a manner that
depends on the PDE describing the gauge).
Though it is difficult
to imagine how any such coordinate transformation (where $\bar{t}$
continues to define
a well-behaved, global time-like foliation) could fundamentally change the
PDE character of the scalar degree of freedom, we have yet to devise
a proof of this.

\subsection{\label{subsec:scaling}Scaling and loss of hyperbolicity}
\begin{figure}[!h] 
\centering
\includegraphics[width=0.97\columnwidth]{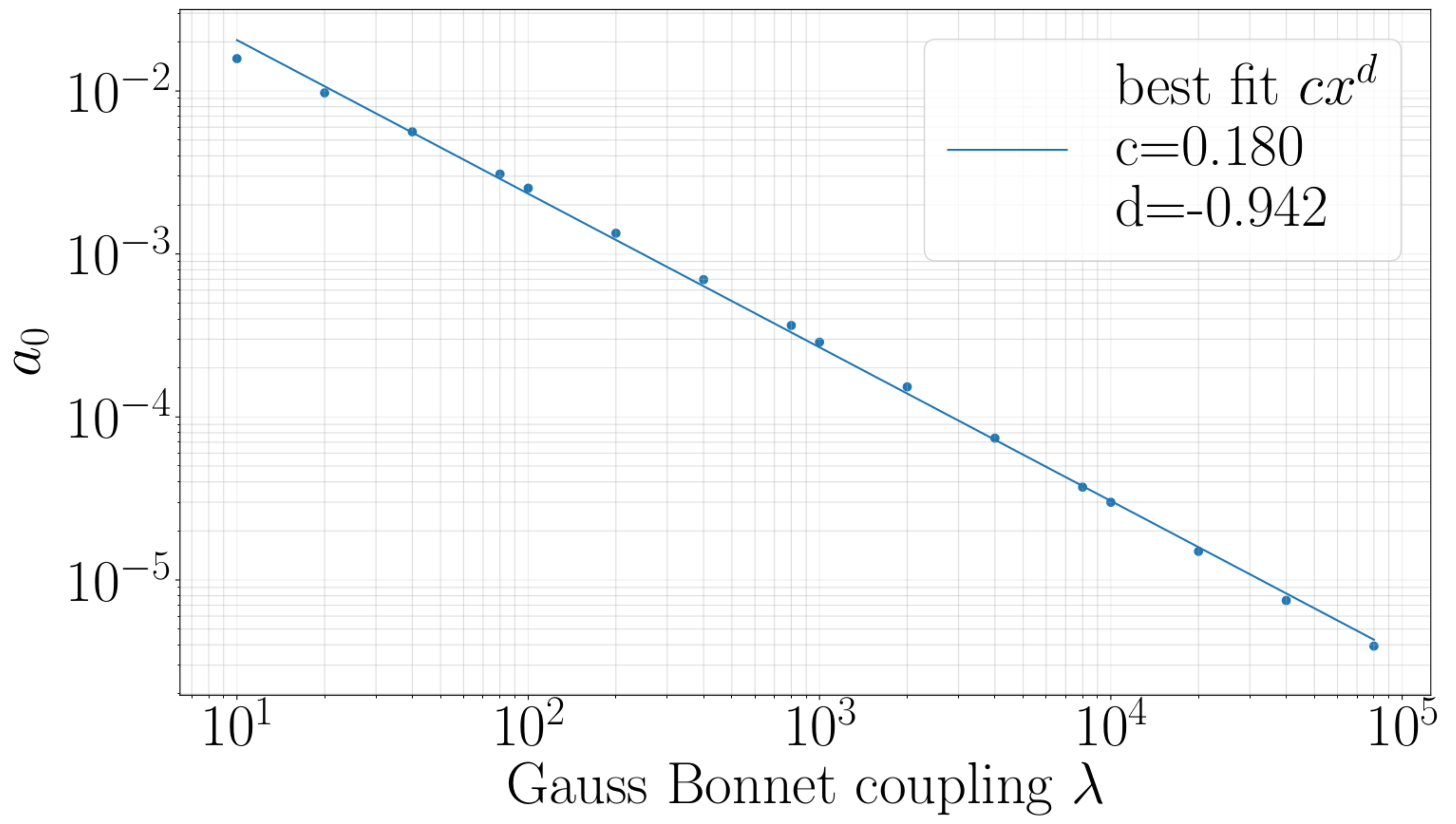}
\caption{
Approximate threshold amplitude for the initial data
Eq.~\eqref{eq:initial_data} (with
fixed $w_0=10,r_0=20$) above which evolution leads to the formation
of a sonic line, as a function of $\lambda$
(run with a spatial resolution $\Delta r\sim1\times10^{-1}$,
and the outer boundary at $R_0=100$).
The ADM mass $m$ scales as $\sim a_0^2$, and is not particularly
sensitive to $\lambda$ in this
range of parameter space, hence the vertical axis also serves
as an indication of the gravitational strength of the initial data
: at $a_0\sim 1.6\times10^{-2}$, $m\sim0.26$, while at 
$a_0\sim 3\times10^{-5}$, $m\sim1\times10^{-6}$. 
}
\label{Fig:scaling_elliptic_region}
\end{figure}

	We now show a result from a survey of evolutions, demonstrating
that the previous example is not a fine-tuned special case within
the initial data 
family Eq.~\eqref{eq:initial_data}, and that formation of an elliptic region
seems to always appear for sufficiently strong coupling, as characterized
by $\eta$.
For this survey we still keep $w_0$ and $r_0$ fixed (now $w_0=10,r_0=20$), but 
for a given $\lambda$ search for the amplitude parameter $a_0$
above which evolution leads to formation of a sonic
line (within the run time of the simulations, corresponding
to roughly a light-crossing time of the domain).
Fig.~\ref{Fig:scaling_elliptic_region} shows the results, and that
the slope of the curve is close to $-1$ suggests
the scaling (\ref{eq:dimensionlessRatio_lambdaLphi}) 
implied by the dimensional analysis does roughly hold in this
set up. 
\section{\label{sec:Conclusion}Conclusion}
We have presented results from a first study of fully non-linear
spherically symmetric gravitational collapse in shift symmetric EdGB 
gravity. We have shown that for the coupling
parameter $\lambda$ sufficiently
large, evolution of certain sets of initial data lead
to situations where the character of the PDEs governing the
dynamical scalar field change from hyperbolic to elliptic in a compact
region of the spacetime. This indicates, within these setups, the
EdGB equations are of mixed type. For sufficiently weak data,
the elliptic region can appear well below the threshold of
BH formation, which might otherwise have censored it
from asymptotic view. This is problematic for the classical theory
to be predictive in the sense of possessing a well-posed Cauchy IVP,
at least for arbitrary values of the coupling parameter.

To gain some intuition for what this implies for 
EdGB gravity serving as a viable modified GR theory to confront with LIGO/Virgo
binary BH merger data, we first note that the smallest
BH solutions that exist in EdGB gravity have a size $L_{SBH}\sim\sqrt{\lambda}$
~\cite{Kanti:1995vq,Sotiriou:2014pfa}. Moreover, as modifications
to GR BH solutions become less pronounced the larger the BH, we do not want
$L_{SBH}\ll O({\rm kms})$ for the theory to remain interesting in this regard.
The scaling relation (\ref{eq:dimensionlessRatio_lambdaLphi}) then
says we will have hyperbolicity issues for compact distributions of scalar
energy on scales $L\lesssim L_{SBH} \sqrt{|\phi_0|}$. This is certainly
problematic if a putative EdGB scalar has an ambient cosmological
background with large density fluctuations. On the other hand,
if we assume the cosmological background for $\phi$ is negligible,
and (away from scalarized BHs) the only significant levels
arise from back-reaction to curvature induced by other matter,
a different scaling relation holds. Using similar dimensional
analysis to that used to derive (\ref{eq:dimensionlessRatio_lambdaLphi}),
one can show that in this case problems arise when the
density $\rho$ of ``ordinary'' matter is greater than $\rho_{SBH}$,
the effective density of the smallest BH allowed. If we then
choose $\rho_{SBH}$ to be slightly greater than nuclear density
(i.e. let the smallest BHs have an effective density
slightly larger than that of neutron stars), and ignoring
problems that might then need to be addressed in the very early
(pre big bang nucleosynthesis) 
universe, our results do not yet rule out EdGB gravity as being
viable and interesting
for stellar mass binary BH mergers.

From the EFT perspective, one could
argue that for $\eta\sim1$ we have entered a regime where
the truncated theory is no longer valid. That is certainly a
sensible stance, but is of little practical use if the higher order
terms or full theory are not known. The alternative then is
to never approach a regime were the small coupling approximation
is violated, but then it is arguable how interesting these theories
can remain as models of beyond-GR modifications in binary BH mergers.

In connection to other work (in addition to that already described
in the introduction), in \cite{Tanahashi:2017kgn} it was
argued that for a wide class of Horndeski theories, including
EdGB gravity, solutions in spherically symmetric may develop shock-like
features in a finite
time. Within our class of initial data (\ref{eq:initial_data}),
below the threshold
of BH formation, we find no evidence for the unbounded growth
of fields or their derivatives
\emph{before} the loss of hyperbolicity of the solution. 

	We re-emphasize that our characteristic analysis is for the spherically
symmetric sector of EdGB gravity, and is tailored to our specific
choice of gauge. In particular, in hyperbolic evolution of EdGB
gravity in a general four dimensional spacetime
there are (expected to be) three, non-gauge, dynamical degrees of freedom. In
a spherically symmetric reduction we can say nothing about the
dynamics or hyperbolicity of the other
two (gravitational) degrees of freedom.
There exist covariant notions of the principal symbol
(e.g. \cite{christodoulou2008mathematical,10.2307/j.ctt1bgzb49.8,
Reall:2014pwa,Papallo:2017qvl}),
from which a covariant notion of hyperbolicity may potentially be derived,
be it in a spherically symmetric reduction or a generic four dimensional
spacetime. 
With regards to mixed type behavior,
the formation of an elliptic region should reduce the
number of dynamical degrees of freedom, which may be apparent from
the perspective of the covariant symbol without the need for
explicit (numerical) solutions. On the other hand, it is
likely that, as we find in the spherically symmetric sector,
loss of hyperbolicity depends on the initial data and coupling
parameter, and so might not be apparent where it occurs without explicit solutions.

For future work, our next step is to continue the study of spherical
collapse in horizon penetrating coordinates, to study under what 
conditions (if any) the BH solutions remain stable and free of mixed type
behavior exterior to the horizons. If viable, we plan to relax
the symmetry restrictions to begin addressing the above equations 
including the gravitational degrees of freedom.

\begin{acknowledgments}
We thank L. Lehner, V. Paschalidis, I. Rodnianski, and M. Taylor
for useful conversations on aspects of 
this project. We thank the anonymous referee for useful comments on
the work. We thank the organizers of the workshop
`Numerical Relativity beyond
General Relativity' and
the Centro de Ciencias de Benasque Pedro Pascual, 2018,
where we completed some of the work presented here.
F.P. acknowledges support from NSF grant PHY-1607449, the Simons
Foundation, and the Canadian Institute For Advanced Research (CIFAR).
Computational resources were provided by the
Feynman cluster at Princeton University.
\end{acknowledgments}

\renewcommand\thefigure{\thesection.\arabic{figure}}   
\setcounter{figure}{0}    

\appendix

\section{\label{sec:NumericalMethods}Numerical methods and convergence}
Here we briefly describe the numerical methods we use, and give one 
convergence result; more details will be given in an upcoming
paper~\cite{RipleyPretorius:2019}.
We solve (\ref{eq_sGB_solved:main}) using second order finite
difference methods. For the results presented here, we implemented
the following
iterative scheme for each time step. In the evolution 
equations for $Q$ and $P$
(\ref{eq_sGB_solved:tDer_Q},\ref{eq_sGB_solved:tDer_P}),
spatial and temporal derivatives are discretized with a Crank-Nicolson stencil.
For each time step we: 
(i) initialize the advanced time level data
at  $t+\Delta t$ for $A,B,P,Q$ with the known solution at time level $t$;
(ii) perform one step of a Newton iteration to solve for a correction
to the unknown values of
 $Q$ and $P$ (\ref{eq_sGB_solved:tDer_Q},\ref{eq_sGB_solved:tDer_P})
at the advanced time level, using a banded matrix solver; 
(iii) integrate the constraints 
(\ref{eq_sGB_solved:rDer_A},\ref{eq_sGB_solved:rDer_B})
using the trapezoidal rule 
to solve for $A$ and $B$ at time $t+\Delta t$, and using latest estimates of 
$P$ and $Q$ from step (ii);
(iv) repeat steps (ii) and (iii) until the residual of the full, non-linear
set of equations~\eqref{eq_sGB_solved:main} is below a tolerance estimated
to be well below truncation error;
(v) apply a Kreiss-Oliger dissipation
filter~\cite{KreissOligerDissipationStandardRef}
to the now known variables at the advanced time.
For initial data at $t=0$, we freely specify $P$ and $Q$
as described above, then solve the constraints as in step (iii).
As a sanity check that the instabilities we observe at late times
are not an artifact of the above time-stepping scheme, we also implemented
several other methods, confirming these results, and which we will
also describe in~\cite{RipleyPretorius:2019}.

	Our simulation has a timelike boundary at a fixed radial distance from
the origin, $r=R_0$. At this outer boundary we imposed 
outgoing Sommerfeld boundary conditions
on the $Q$ and $P$ fields,
$\partial_t(rQ) + \partial_r(rQ)=0$,
$\partial_t(rP) + \partial_r(rP)=0$. As the $A$ and $B$ fields
are determined from the first order ordinary differential equations
Eqs~\eqref{eq_sGB_solved:rDer_A} and \eqref{eq_sGB_solved:rDer_B},
 imposing regularity at the origin
specifies their behavior over the entire computational domain.
	Imposing regularity at the origin, where polar
coordinates are singular, dictates
$	\partial_rA|_{r=0} 
	= \partial_rB|_{r=0} 
	= B|_{r=0} 
	= Q|_{r=0} 
	= \partial_rP|_{r=0} = 0
$.
The condition $\partial_rB|_{r=0} = 0$ is automatically enforced
by $B|_{r=0}
	= Q|_{r=0} 
	= \partial_rP|_{r=0} = 0
$ and (\ref{eq_sGB_solved:rDer_B}).
The condition $\partial_rA|_{r=0} = 0$ is automatically enforced
by $B|_{r=0}=\partial_rB|_{r=0}
	= Q|_{r=0} 
	= \partial_rP|_{r=0} = 0
$,
and (\ref{eq_sGB_solved:rDer_A}).
With the coordinates \eqref{eq:polar_coordinates} 
we have the residual gauge
freedom to linearly shift $A$ by a function $f(t)$;
we use this to set $A(t,R_0)=0$.

\begin{figure}[h] 
\begin{center}
\includegraphics[width=1.0\columnwidth]{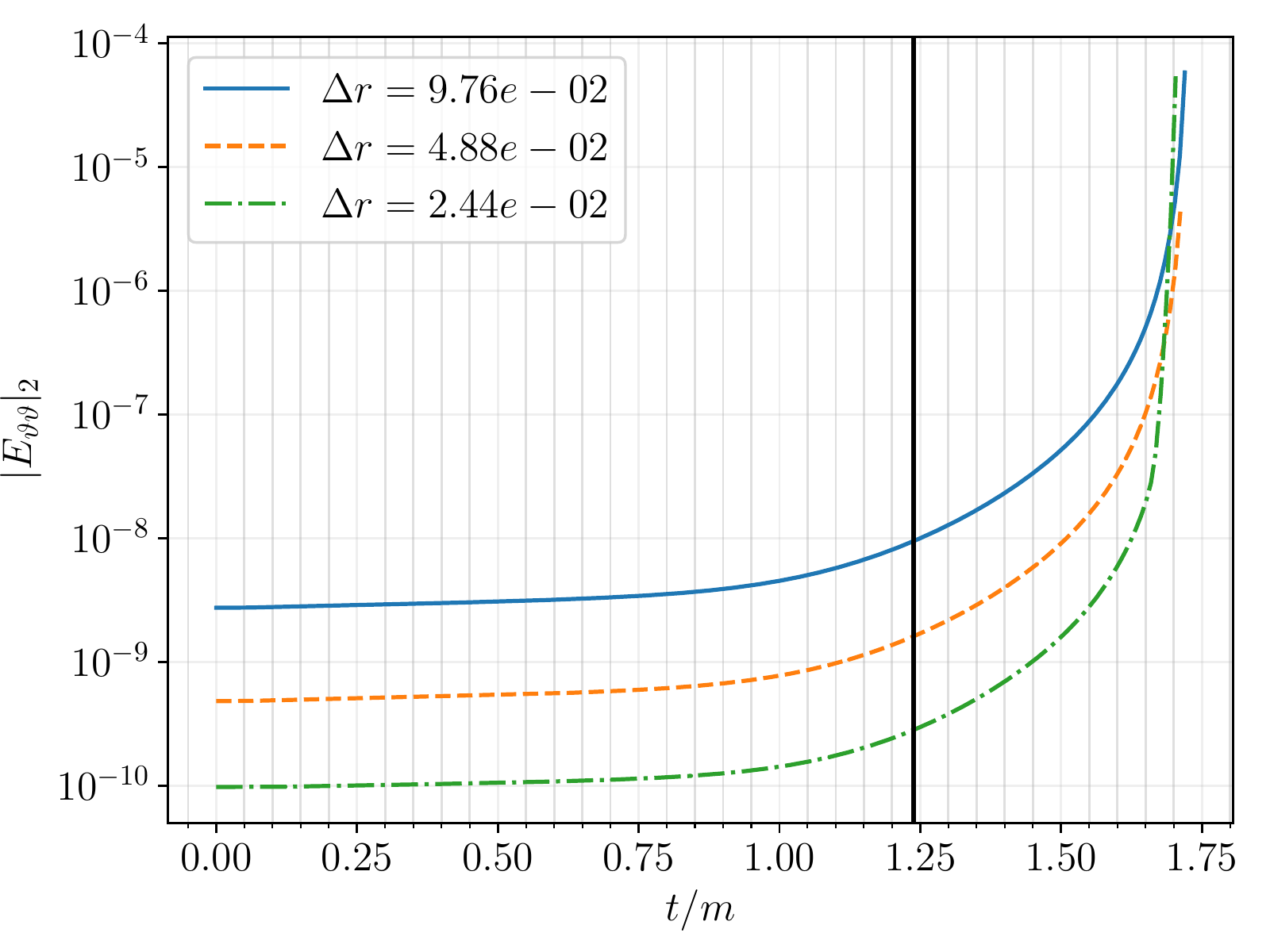}
\caption{
\label{Fig:convergence}
The $L_2$ norm of the $\vartheta\vartheta$ component of
Eq.~\eqref{eq_EinsteinEquations:metriceom}. 
We begin 
to lose convergence after the formation of the elliptic region
(black vertical line; compare with Fig.~\ref{fig:characteristics}). 
}
\end{center}
\end{figure}

Fig.~\ref{Fig:convergence} shows the $\vartheta\vartheta$ component
of the equations of motion for the example described in this paper; we
see second order convergence to zero up to and slightly
past the formation of the elliptic region. This behavior is typical
of the independent residuals for all the different solution techniques we
tried. As discussed in the main article,
after the formation of the elliptic region,
we expect the system of equations solved as a system of hyperbolic PDEs to become
ill posed. Short wavelength modes should then grow exponentially,
and the fastest growing modes will have wavelengths proportional to the mesh
spacing, hence convergence will be lost, and higher resolution simulations
will begin to ``crash'' sooner. What complicates estimating
the exact time of blow-up, is since our initial data is smooth,
the amplitudes of the short wavelength modes that become
unstable in the elliptic region are mostly sourced by truncation
error, and so will initially be smaller for higher resolution.
In qualitative agreement with these expectations,
in Fig.~\ref{Fig:convergence} we observe 
that the independent residuals of the higher resolution simulations 
grow more quickly than they do for the lower resolution simulations
after the formation of the elliptic region. 

\bibliography{TEdGB_polarCoordinate_shorterVersion_01}

\end{document}